\documentclass{article}

\PassOptionsToPackage{numbers, compress}{natbib}
\usepackage[final]{neurips_2022_ml4ps}

\usepackage[utf8]{inputenc} % allow utf-8 input
\usepackage[T1]{fontenc}    % use 8-bit T1 fonts
\usepackage{url}            % simple URL typesetting
\usepackage{booktabs}       % professional-quality tables
\usepackage{amsfonts}       % blackboard math symbols
\usepackage{nicefrac}       % compact symbols for 1/2, etc.
\usepackage{microtype}      % microtypography
\usepackage{xcolor}         % colors

\usepackage{bm}             % bold maths
\usepackage{amsmath}
\usepackage{amssymb}
\usepackage[pdftex]{graphicx}

\newcommand{\blank}{\mathrel{\;\cdot\;}}

\newcommand\corestr[2]{{
  \left.\kern-\nulldelimiterspace 
  #1
  \vphantom{\big|}
  \right|^{#2}
}}

\title{Physics-Informed CNNs for Super-Resolution of Sparse Observations on Dynamical Systems}

\author{
  Daniel Kelshaw \\
  Department of Aeronautics \\
  Imperial College London \\
  \texttt{djk21@imperial.ac.uk} \\
  \And        
  Georgios Rigas \\
  Department of Aeronautics \\
  Imperial College London \\
  \texttt{g.rigas@imperial.ac.uk} \\
  \And
  Luca Magri \\
  Department of Aeronautics \\
  Imperial College London, \\
  Alan Turing Institute \\
  \texttt{l.magri@imperial.ac.uk} \\
}

\begin{document}

\maketitle

\begin{abstract}
    In the absence of high-resolution samples, super-resolution of sparse observations 
    on dynamical systems is a challenging problem with wide-reaching applications in experimental
    settings. We showcase the application of physics-informed convolutional neural networks for
    super-resolution of sparse observations on grids. Results are shown for the chaotic-turbulent
    Kolmogorov flow, demonstrating the potential of this method for resolving finer scales of 
    turbulence when compared with classic interpolation methods, and thus effectively 
    reconstructing missing physics.
\end{abstract}

% Providing snapshot of the results
\begin{figure}[ht]
    \centering
    \includegraphics[width=4.5in, height=1.4in, keepaspectratio]{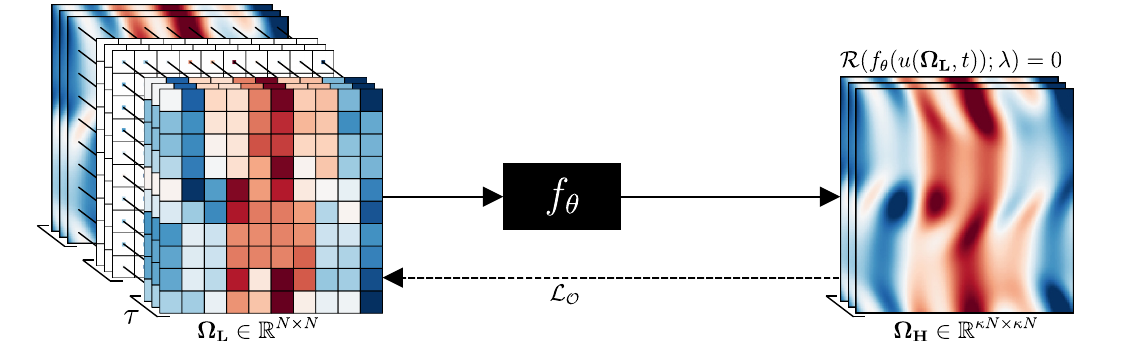}
    \caption{Diagrammatic overview of the super-resolution task.}
    \label{fig:overview}
\end{figure}

\section{Introduction}
The application of machine learning for dynamical systems is gaining traction, providing means to extract physical information 
from data, and reducing the dependence on running computationally expensive simulations \citep{Brunton2020}. In many cases, access
to only sparse or partial observations of a dynamical system is a limiting factor, obscuring the underlying dynamics and providing 
a challenge for system identification \cite{Brunton2016}. Super-resolution methods offer the means for high-fidelity state 
reconstruction from limited observations, a problem of fundamental importance in the physical sciences.

Convolutional neural networks (CNNs) are prominent in the domain of image reconstruction due to their inherent ability to exploit
spatial correlations \citep{Dong2014, Shi2016, Yang2019}. For the classic data-driven approach, there is a dependency on access 
to samples of high-resolution data. In the absence of ground-truth labels, as with observations on a dynamical system, a common 
approach is to impose prior knowledge of the physics, regularising predictions with respect to known governing equations \citep{Lagaris1998}. 
The introduction of physics-informed neural networks (PINNs) has provided new tools for physically-motivated problems, exploiting 
the automatic-differentiation paradigm provided by neural networks to constrain gradients of the physical system 
\cite{Raissi2019, Cranmer2020, Cai2021}. Applications of PINNs for super-resolution show promising results for simple systems, 
using sparse observations for accurate reconstruction of high-resolution fields \citep{Eivazi2022}.
% Super-resolution using PINNs shows promising results for simple systems, using sparse 
% observations for accurate reconstruction of high-resolution fields \citep{Eivazi2022}.

Applications of physics-informed CNNs for super-resolution are less prevalent. While \citet{Liu2020} demonstrate a CNN for 
super-resolution of observations on a dynamical system, there is a strict dependence on high-resolution examples to train the model. 
\citet{Gao2021} explore physics-informed CNNs for the super-resolution of a steady flow field, producing a one-to-one
mapping for stationary solutions of the Navier-Stokes equations. Physics-based regularisation of the steady solution neglects any temporal 
component, introducing further complexities when considering dynamical systems.

Our work extends on the current use of physics-informed CNNs for super-resolution, considering the application to dynamical systems 
where the temporal nature cannot be neglected. We showcase the super-resolution of sparse, spatial observations of the chaotic-turbulent
Kolmogorov flow, highlighting the ability to recover finer scales of turbulence through reconstruction of the missing physics.

\section{Defining the Dynamical System} \label{sec:methodology}

% ## Introduction ####################################################################################################################
Our work considers super-resolution of sparse observations on dynamical systems of the form

\begin{equation} \label{eqn:dynamical_system}
    \partial_t \bm{u} - \mathcal{N}(\bm{u}; \lambda) = 0
        \qquad
        \text{with}
            \quad \bm{u} = u(\bm{x}, t),
            \quad \bm{x} \in \Omega \subset \mathbb{R}^{n}, 
            \quad t \in [0, T] \subset \mathbb{R}^{+},
\end{equation}

where $u: \Omega \times [0, T] \rightarrow \mathbb{R}^{n}$, $\mathcal{N}$ is a sufficiently smooth differential operator, and $\lambda$ 
are the physical parameters of the system. We define the residual of the system as the left-hand side of equation (\ref{eqn:dynamical_system})

\begin{equation}
    \mathcal{R}(\bm{u}; \lambda) \triangleq \partial_t \bm{u} - \mathcal{N}(\bm{u}; \lambda),
\end{equation}

such that $\mathcal{R}(\bm{u}; \lambda) = 0$ when $u(\bm{x}, t)$ is a solution to the partial differential equation (PDE).

% ## Super Resolution Task ###########################################################################################################

\section{The Super-Resolution Task} \label{sec:task}
Given sparse observations on a low-resolution grid, we aim to reconstruct the underlying solution to the PDE on a high-resolution grid. 
Mathematically, we denote this process by the mapping

\begin{equation}
    f_{\theta}: u(\bm{\Omega_{L}}, t) \rightarrow u(\bm{\Omega_{H}}, t),
\end{equation}

where the domain $\Omega$ is discretised on uniform, structured grids $\bm{\Omega_{L}} \subset \mathbb{R}^{N^n}$, $\bm{\Omega_{H}} \subset \mathbb{R}^{M^n}$
such that $\bm{\Omega_{L}} \cap \bm{\Omega_{H}} = \bm{\Omega_{L}}$, and $M = \kappa N$ where $\kappa \in \mathbb{N}^{+}$ is the up-sampling factor. An overview
is provided in Figure \ref{fig:overview}. We further discretise the time-domain, providing $\mathcal{T} = \{ t_i \in [0, T] \}_{i=0}^{N_t}$ for $N_t$ samples. 
Approximating the mapping $f_{\theta}$ as a CNN, we optimise weights $\theta$ of the network to minimise the loss

\begin{equation} \label{eqn:total_loss}
    \mathcal{L}_{\theta} = \alpha \mathcal{L}_{\mathcal{O}} + \mathcal{L}_{\mathcal{R}},
\end{equation}

where $\alpha$ represents a fixed, empirical weighting parameter. We further define each loss term as

\begin{equation} \label{eqn:lo_lr}
\begin{aligned}[c]
    \mathcal{L}_{\mathcal{O}} &= \frac{1}{\lvert \mathcal{T} \rvert} \sum_{t \in \mathcal{T}} \lVert 
        \corestr{f_{\theta}(u(\bm{\Omega_{L}}, t))}{\bm{\Omega_{L}}} - u(\bm{\Omega_{L}}, t)
    \rVert_{\bm{\Omega_{L}}}^{2}, \\
    \mathcal{L}_{\mathcal{R}} &= \frac{1}{\lvert \mathcal{T} \rvert} \sum_{t \in \mathcal{T}} \lVert 
       \mathcal{R}(f_{\theta}(u(\bm{\Omega_{L}}, t)); \lambda)
    \rVert_{\bm{\Omega_{H}}}^{2},
\end{aligned}
\end{equation}

% ## WHAT DOES EACH LOSS DO ##########################################################################################################
where $\corestr{f_{\theta}(\blank)}{\bm{\Omega_{L}}}$ denotes the corestriction of $\bm{\Omega_{H}}$ on $\bm{\Omega_{L}}$, and $\lVert \blank \rVert_{\Omega}$ 
represents the $\ell^2$-norm over the given domain. The observation-based loss $\mathcal{L}_{\mathcal{O}}$ seeks to minimise the distance
between known observations and their corresponding predictions - not accounting for additional high-resolution information unavailable 
to the system. We regularise network predictions with the residual-based term $\mathcal{L}_{\mathcal{R}}$, seeking to ensure that 
realisations of high-resolution fields satisfy the governing PDEs. As a consequence of imposing prior knowledge of the governing equations
through the residual-based loss term, we are effectively able to condition the underlying physics on the observed data, allowing us to
recover the underlying solution on the grid $\bm{\Omega_{H}}$.\footnote{All code is available on GitHub: \url{https://github.com/magrilab/PISR}}

\section{Results} \label{sec:results} 

% ## INTRODUCING THE KOLMOGOROV FLOW #################################################################################################
We consider super-resolution for the Kolmogorov flow, a solution of the incompressible Navier-Stokes equations. The system is evaluated 
on the domain $\Omega \in [0, 2\pi) \subset \mathbb{R}^{2}$ with periodic boundary conditions applied on $\partial \Omega$, and a stationary, 
spatially-varying sinusoidal forcing term $g(\bm{x})$. The Kolmogorov flow is prominent in studies of turbulence, providing
a suitable case study for the super-resolution task; this allows us to evaluate the quality of predictions with respect to the turbulent
energy cascade \citep{Fylladitakis2018}. The standard continuity and momentum equations are

\begin{align} \label{eqn:kolmogorov}
\begin{split}
    \nabla \cdot \bm{u} &= 0, \\
    \partial_t \bm{u} + \bm{u} \cdot \nabla \bm{u} &= - \nabla p + \nu \Delta \bm{u} + g(\bm{x}),
\end{split}
\end{align}

where $p, \nu$ denote the scalar pressure field and kinematic viscosity respectively. We take $\nu = \nicefrac{1}{34}$ to ensure 
chaotic-turbulent dynamics, and prescribe the standard forcing $g(\bm{x}) = [\sin(4\bm{x}_{2}), 0]^{\top}$.

% ## SOLVER DETAILS ##################################################################################################################
\subsection{Differentiable Pseudospectral Discretisation} \label{sec:solver}
We utilise a differentiable pseudospectral spatial discretisation for the Kolmogorov flow, enabling backpropagation for the residual-based
loss, $\mathcal{L}_{\mathcal{R}}$. By eliminating the pressure term, our discretisation handles the continuity constraint implicitly, allowing
us to neglect the term in the loss \cite{Canuto1988}. Data is generated by time-integration of the dynamical system with the forward-Euler
scheme, taking a time-step $\Delta t = 0.005$ to ensure numerical stability according to the Courant–Friedrichs–Lewy condition. We evaluate
the residual-based loss $\mathcal{L}_{\mathcal{R}}$ in the Fourier domain, re-defining the loss

\begin{equation}
    \mathcal{L}_{\mathcal{R}} = \frac{1}{\lvert \mathcal{T} \rvert} \sum_{t \in \mathcal{T}} \lVert
        \partial_t \hat{f}_{\theta}(u(\bm{\Omega_{L}}, t)) - \hat{\mathcal{N}}(\hat{f}_{\theta}(u(\bm{\Omega_{L}}, t)))
    \rVert_{\bm{\hat{\Omega}_{k}}}^{2},
\end{equation}

where $\bm{\hat{\Omega}_{k}} \in \mathbb{Z}^{K^n}$ is the discretised Fourier wavespace grid, $\hat{f}_{\theta} = \mathcal{F} \circ f_{\theta}$
where $\mathcal{F}$ is the Fourier operator, and $\hat{\mathcal{N}}$ denotes the Fourier differential operator; represented by the 
differentiable discretisation. We consider evaluation of the loss over discrete time-windows, each consisting of $\tau \geq 2$ 
consecutive time-steps. This windowing procedure allows for localised computation of the residual in the time domain, alleviating the 
requirement for the observations to be totally contiguous in time.

% ## EXPERIMENT SETUP ################################################################################################################
\subsection{Numerical Experiments} \label{sec:numerical_experiments}
We solve the flow in the Fourier domain on a grid $\bm{\hat{\Omega}_{k}} \in \mathbb{Z}^{61 \times 61}$. Data is generated on the
high-resolution grid $\bm{\Omega_{H}} \in \mathbb{R}^{150 \times 150}$ prior to extracting the low-resolution grid $\bm{\Omega_{L}} \in \mathbb{R}^{10 \times 10}$
of observations. A total of $2048$ time-windows are used for training, with a further $256$ for validation; taking $\tau = 2$ in each instance. 
We employ the VDSR architecture \cite{Kim2016}, a variation on the common VGG-net \citep{Simonyan15}. By prepending the convolutional layers 
with a bi-cubic upsampling layer, we reformulate the problem in-terms of learning the residual between the interpolation, and the high-resolution
field; exploiting benefits associated with residual learning problems \citep{Li2018}. Periodic boundary conditions are embedded in the model 
through the use of periodic padding in the convolutional layers. For training we use the Adam optimizer with a learning rate of $3\times10^{-4}$, 
weighting the loss with $\alpha = 1\times10^5$.

\begin{figure}[ht]
    \centering
    \includegraphics[width=\linewidth]{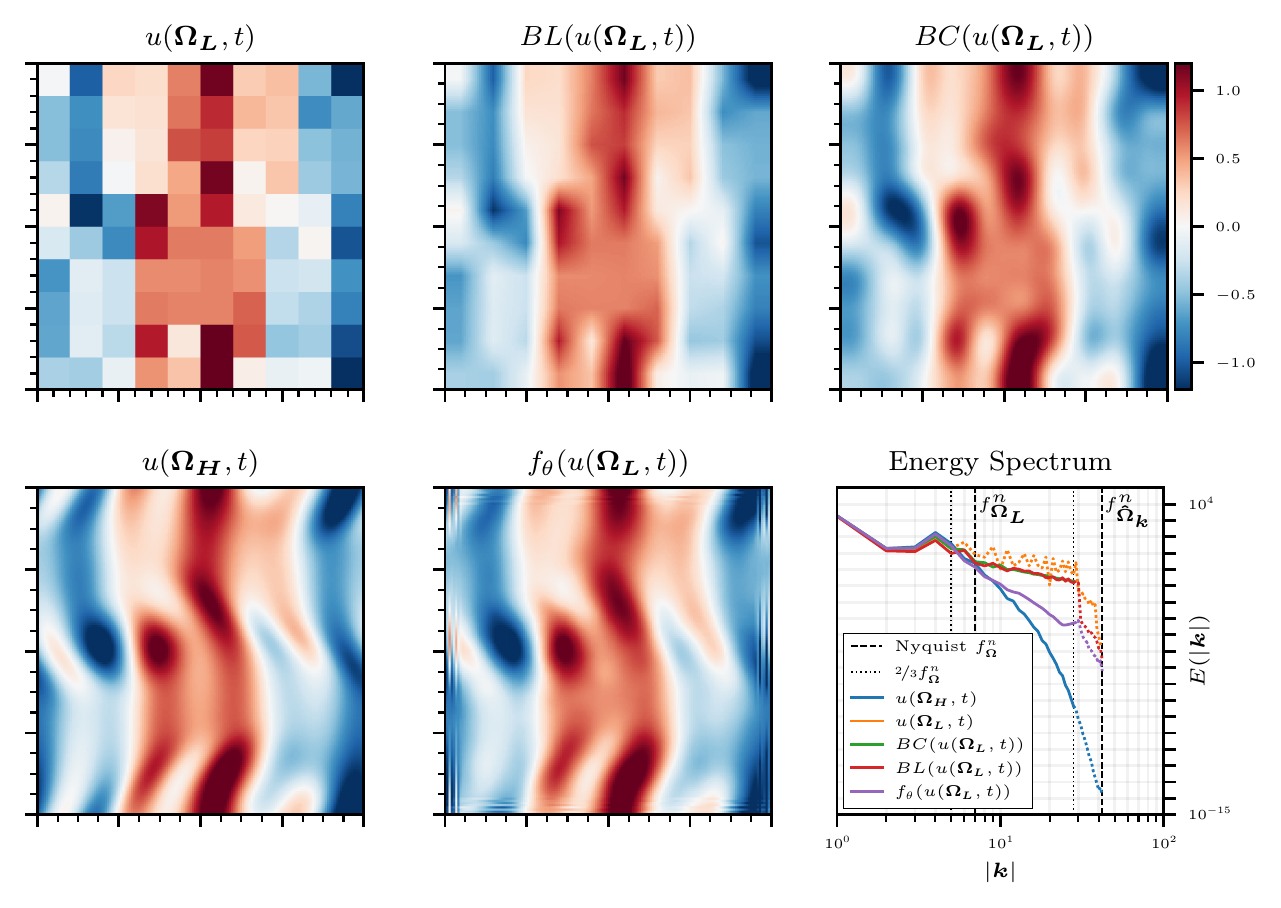}
    \caption{
        Super-resolution of $u(\bm{\Omega_{L}}, t) \in \mathbb{R}^{10\times10}$ to $u(\bm{\Omega_{H}}, t) \in \mathbb{R}^{150\times150}$, demonstrating: 
        bi-linear interpolation, $BL(u(\bm{\Omega_{L}}, t))$; bi-cubic interpolation, $BC(u(\bm{\Omega_{L}}, t))$; and predictions, $f_{\theta}(u(\bm{\Omega_{L}}, t))$.
    }
    \label{fig:snapshot}
\end{figure}

Results are compared with bi-linear ($BL$) and bi-cubic ($BC$) interpolation, measuring performance by the relative $\ell^2$-error between network
predictions $f_{\theta}(\bm{\Omega_{L}}, t)$, and the true high-resolution fields, $u(\bm{\Omega_{H}}, t)$. Figure \ref{fig:snapshot} shows a snapshot of the 
results, with network predictions yielding qualitatively superior results when compared with the interpolated alternatives. Quantitatively we observe
a lower error for the network predictions, achieving a relative $\ell^2$-error of $0.0872$, compared with $0.2091$ for bi-linear interpolation and 
$0.1717$ for bi-cubic interpolation; averaging results over available data.\footnote{All experiments were run on a single NVIDIA Quadro RTX 8000.} 

% Providing snapshot of the results

The energy spectrum is characteristic of the direct energy cascade observed in turbulent flows, a multi-scale phenomenon which sees energy content
decaying with increasing wavenumber. In Figure \ref{fig:snapshot}, we observe that the energy content of the low-resolution field diverges from that
of the high-resolution field; a consequence of spectral aliasing. Applying the $\nicefrac{2}{3}$ de-aliasing rule to the Nyquist frequencies, we see
that aliasing is introduced at  $\lvert \bm{k} \rvert = 5, 28$ for the low- and high-resolution fields respectively. This correlates with the divergence in
energy content observed in the spectrum. We find that fields super-resolved by the network are capable of capturing finer scales of turbulence compared
to low-resolution and interpolation approaches, prior to diverging at $\lvert \bm{k} \rvert = 10^{1}$. This result is indicative of the impact of the 
residual-based loss, signifying the ability of the network to act beyond simple interpolation. The network is capable of de-aliasing, or inferring 
missing physics.

Despite the use of periodic padding in the convolutional layers, we observe artefacts on the boundaries of the predicted field. These artefacts are 
largely responsible for the divergence of the predicted energy spectrum from the true spectrum and provide a consideration for future work.

\section{Conclusion} \label{sec:conclusion}

% ## CONCLUDING REMARKS ##############################################################################################################
In this work, we have demonstrated the application of physics-informed CNNs for the super-resolution of the chaotic-turbulent Kolmogorov
flow. Our results show improved performance when compared with standard interpolation methods. Evaluation of the turbulent energy spectrum
highlights the ability of the network to resolve turbulent structures on a finer scale than available in the sparse observations. We note
that this reconstruction of turbulence showcases the ability to resolve missing physics in the sparse observations, a consequence
of embedding prior knowledge of the physics in the loss term. With a view on the spectral representation, we find that the network 
effectively mitigates the aliasing introduced by taking sparse observations. This work opens opportunities for the accurate reconstruction of 
dynamical systems from sparse observations, as is often the case in experimental settings for the physical sciences.

\acksection{
D. Kelshaw. and L. Magri. acknowledge support from the UK EPSRC. 
L. Magri gratefully acknowledges financial support from the ERC Starting Grant PhyCo 949388.
}

% \newpage
% % ## IMPACT STATEMENT ################################################################################################################
% \section{Impact Statement}
% Experimental settings are prevalent in the physical sciences. Although we strive to capture information across the entire domain, we are
% often restricted to taking sparse observations due to limitations imposed by available measurement techniques; for example: probes at 
% discrete locations in a fluid flow. These sparse observations constitute a bottleneck for scientific discovery, making detailed analysis
% of experimental results intractable. Our work opens opportunities for recovering information across the entire domain by imposing knowledge
% of the underlying physics. We believe that the methods described in this work will allow for more efficient analysis in experimental 
% settings, and help to alleviate the burden placed on high-fidelity computational methods used in lieu of adequate observations. In addition,
% we note that the use of physics-informed methods reduces the dependence on data, providing tools which allow for more efficient computation
% by imposing prior knowledge. We view that our contributions can help accelerate scientific discovery, and reduce energy consumption from
% unnecessary simulations - all of which provide positive societal impact.

% % ####################################################################################################################################

\bibliography{references}

% Generated by IEEEtranN.bst, version: 1.14 (2015/08/26)
\begin{thebibliography}{17}
\providecommand{\natexlab}[1]{#1}
\providecommand{\url}[1]{#1}
\csname url@samestyle\endcsname
\providecommand{\newblock}{\relax}
\providecommand{\bibinfo}[2]{#2}
\providecommand{\BIBentrySTDinterwordspacing}{\spaceskip=0pt\relax}
\providecommand{\BIBentryALTinterwordstretchfactor}{4}
\providecommand{\BIBentryALTinterwordspacing}{\spaceskip=\fontdimen2\font plus
\BIBentryALTinterwordstretchfactor\fontdimen3\font minus
  \fontdimen4\font\relax}
\providecommand{\BIBforeignlanguage}[2]{{%
\expandafter\ifx\csname l@#1\endcsname\relax
\typeout{** WARNING: IEEEtranN.bst: No hyphenation pattern has been}%
\typeout{** loaded for the language `#1'. Using the pattern for}%
\typeout{** the default language instead.}%
\else
\language=\csname l@#1\endcsname
\fi
#2}}
\providecommand{\BIBdecl}{\relax}
\BIBdecl

\bibitem[Brunton et~al.(2020)Brunton, Noack, and Koumoutsakos]{Brunton2020}
S.~Brunton, B.~Noack, and P.~Koumoutsakos, ``Machine learning for fluid
  mechanics,'' \emph{Annual Review of Fluid Mechanics}, vol.~52, pp. 477--508,
  1 2020.

\bibitem[Brunton et~al.(2016)Brunton, Proctor, and Kutz]{Brunton2016}
S.~L. Brunton, J.~L. Proctor, and J.~N. Kutz, ``Discovering governing equations
  from data by sparse identification of nonlinear dynamical systems,''
  \emph{Proceedings of the National Academy of Sciences}, vol. 113, no.~15, pp.
  3932--3937, 2016.

\bibitem[Dong et~al.(2014)Dong, Loy, He, and Tang]{Dong2014}
C.~Dong, C.~C. Loy, K.~He, and X.~Tang, ``Learning a deep convolutional network
  for image super-resolution,'' in \emph{European Conference on Computer
  Vision}, 2014, pp. 184--199.

\bibitem[Shi et~al.(2016)Shi, Caballero, Huszár, Totz, Aitken, Bishop,
  Rueckert, and Wang]{Shi2016}
W.~Shi, J.~Caballero, F.~Huszár, J.~Totz, A.~P. Aitken, R.~Bishop,
  D.~Rueckert, and Z.~Wang, ``Real-time single image and video super-resolution
  using an efficient sub-pixel convolutional neural network,''
  \emph{Proceedings of the IEEE Conference on Computer Vision and Pattern
  Recognition (CVPR)}, 9 2016.

\bibitem[Yang et~al.(2019)Yang, Zhang, Tian, Wang, Xue, and Liao]{Yang2019}
W.~Yang, X.~Zhang, Y.~Tian, W.~Wang, J.-H. Xue, and Q.~Liao, ``Deep learning
  for single image super-resolution: A brief review,'' \emph{IEEE Transactions
  on Multimedia}, vol.~21, pp. 3106--3121, 12 2019.

\bibitem[Lagaris et~al.(1998)Lagaris, Likas, and Fotiadis]{Lagaris1998}
I.~E. Lagaris, A.~Likas, and D.~I. Fotiadis, ``Artificial neural networks for
  solving ordinary and partial differential equations,'' \emph{IEEE
  Transactions on Neural Networks}, vol.~9, pp. 987--1000, 1998.

\bibitem[Raissi et~al.(2019)Raissi, Perdikaris, and Karniadakis]{Raissi2019}
M.~Raissi, P.~Perdikaris, and G.~E. Karniadakis, ``Physics-informed neural
  networks: A deep learning framework for solving forward and inverse problems
  involving nonlinear partial differential equations,'' \emph{Journal of
  Computational Physics}, vol. 378, pp. 686--707, 2 2019.

\bibitem[Cranmer et~al.(2019)Cranmer, Greydanus, Hoyer, Battaglia, Spergel, and
  Ho]{Cranmer2020}
M.~Cranmer, S.~Greydanus, S.~Hoyer, P.~Battaglia, D.~Spergel, and S.~Ho,
  ``Lagrangian neural networks,'' in \emph{ICLR 2020 Workshop on Integration of
  Deep Neural Models and Differential Equations}, 2019.

\bibitem[Cai et~al.(2021)Cai, Mao, Wang, Yin, and Karniadakis]{Cai2021}
S.~Cai, Z.~Mao, Z.~Wang, M.~Yin, and G.~E. Karniadakis, ``Physics-informed
  neural networks (pinns) for fluid mechanics: a review,'' \emph{Acta Mechanica
  Sinica}, vol.~37, pp. 1727--1738, 12 2021.

\bibitem[Eivazi and Vinuesa(2022)]{Eivazi2022}
\BIBentryALTinterwordspacing
H.~Eivazi and R.~Vinuesa, ``Physics-informed deep-learning applications to
  experimental fluid mechanics,'' 2022. [Online]. Available:
  \url{https://arxiv.org/abs/2203.15402}
\BIBentrySTDinterwordspacing

\bibitem[Liu et~al.(2020)Liu, Tang, Huang, and Lu]{Liu2020}
B.~Liu, J.~Tang, H.~Huang, and X.-Y. Lu, ``Deep learning methods for
  super-resolution reconstruction of turbulent flows,'' \emph{Physics of
  Fluids}, vol.~32, p. 25105, 2020.

\bibitem[Gao et~al.(2021)Gao, Sun, and Wang]{Gao2021}
H.~Gao, L.~Sun, and J.-X. Wang, ``Super-resolution and denoising of fluid flow
  using physics-informed convolutional neural networks without high-resolution
  labels,'' \emph{Physics of Fluids}, vol.~33, p. 073603, 7 2021.

\bibitem[Fylladitakis(2018)]{Fylladitakis2018}
E.~D. Fylladitakis, ``Kolmogorov flow: Seven decades of history,''
  \emph{Journal of Applied Mathematics and Physics}, vol.~6, pp. 2227--2263,
  2018.

\bibitem[Canuto et~al.(1988)Canuto, Hussaini, Quarteroni, and Zang]{Canuto1988}
C.~Canuto, M.~Y. Hussaini, A.~Quarteroni, and T.~A. Zang, \emph{Spectral
  Methods in Fluid Dynamics}.\hskip 1em plus 0.5em minus 0.4em\relax Springer
  Berlin Heidelberg, 1988.

\bibitem[Kim et~al.(2016)Kim, Lee, and Lee]{Kim2016}
J.~Kim, J.~K. Lee, and K.~M. Lee, ``Accurate image super-resolution using very
  deep convolutional networks,'' in \emph{Proceedings of the IEEE Conference on
  Computer Vision and Pattern Recognition (CVPR)}, 2016.

\bibitem[Simonyan and Zisserman(2015)]{Simonyan15}
K.~Simonyan and A.~Zisserman, ``Very deep convolutional networks for
  large-scale image recognition,'' in \emph{International Conference on
  Learning Representations}, 2015.

\bibitem[Li et~al.(2018)Li, Xu, Taylor, Studer, and Goldstein]{Li2018}
H.~Li, Z.~Xu, G.~Taylor, C.~Studer, and T.~Goldstein, ``Visualizing the loss
  landscape of neural nets,'' in \emph{Neural Information Processing Systems},
  2018, pp. 6391--6401.

\end{thebibliography}

\end{document}